\documentclass[preprint,5p,sort&compress,longtitle]{elsarticle}  

\usepackage{graphicx}  
\usepackage{dcolumn}   
\usepackage{bm}        
\usepackage{amssymb}   
\usepackage{textcomp}
\usepackage{verbatim}
\usepackage{amsmath}
\usepackage{appendix}
\usepackage{enumitem}
\usepackage{epstopdf}
\usepackage{subfig}
\usepackage[utf8]{inputenc}  
\usepackage[draft]{hyperref}

\providecommand*{\unit}[1]{\ensuremath{\mathrm{\,#1}}}
\def\1/2{\frac{1}{2}}
\def\3/2{\frac{3}{2}}

\def\para{\parallel}

\begin{document}

\begin{frontmatter}

\title{Photon beam asymmetry $\Sigma$ for $\eta$ and $\eta^\prime$ photoproduction from the proton}

\newcommand*{\ANL}{Argonne National Laboratory, Argonne, Illinois 60439}
\newcommand*{\ANLindex}{1}
\newcommand*{\ASU}{Arizona State University, Tempe, Arizona 85287-1504}
\newcommand*{\ASUindex}{2}
\newcommand*{\BONNB}{Bethe Center for Theoretical Physics, Universit\"{a}t Bonn, Germany}
\newcommand*{\BONNBindex}{3}
\newcommand*{\BONNHI}{Helmholtz-Institut f\"{u}r Strahlen- und Kernphysik, Universit\"{a}t Bonn, Germany}
\newcommand*{\BONNHIindex}{4}
\newcommand*{\CSUDH}{California State University, Dominguez Hills, Carson, CA 90747}
\newcommand*{\CSUDHindex}{5}
\newcommand*{\CMU}{Carnegie Mellon University, Pittsburgh, Pennsylvania 15213}
\newcommand*{\CMUindex}{6}
\newcommand*{\CUA}{Catholic University of America, Washington, D.C. 20064}
\newcommand*{\CUAindex}{7}
\newcommand*{\SACLAY}{Irfu/SPhN, CEA, Universit\'e Paris-Saclay, 91191 Gif-sur-Yvette, France}
\newcommand*{\SACLAYindex}{8}
\newcommand*{\CNU}{Christopher Newport University, Newport News, Virginia 23606}
\newcommand*{\CNUindex}{9}
\newcommand*{\UCONN}{University of Connecticut, Storrs, Connecticut 06269}
\newcommand*{\UCONNindex}{10}
\newcommand*{\FU}{Fairfield University, Fairfield CT 06824}
\newcommand*{\FUindex}{11}
\newcommand*{\FERRARAU}{Universita' di Ferrara , 44121 Ferrara, Italy}
\newcommand*{\FERRARAUindex}{12}
\newcommand*{\FIU}{Florida International University, Miami, Florida 33199}
\newcommand*{\FIUindex}{13}
\newcommand*{\FSU}{Florida State University, Tallahassee, Florida 32306}
\newcommand*{\FSUindex}{14}
\newcommand*{\GATCHINA}{NRC ``Kurchatov'' Institute, PNPI, Gatchina 188300, Russia}
\newcommand*{\GATCHINAindex}{15}
\newcommand*{\Genova}{Universit$\grave{a}$ di Genova, 16146 Genova, Italy}
\newcommand*{\Genovaindex}{16}
\newcommand*{\GWUI}{The George Washington University, Washington, DC 20052}
\newcommand*{\GWUIindex}{17}
\newcommand*{\ISU}{Idaho State University, Pocatello, Idaho 83209}
\newcommand*{\ISUindex}{18}
\newcommand*{\INFNFE}{INFN, Sezione di Ferrara, 44100 Ferrara, Italy}
\newcommand*{\INFNFEindex}{19}
\newcommand*{\INFNFR}{INFN, Laboratori Nazionali di Frascati, 00044 Frascati, Italy}
\newcommand*{\INFNFRindex}{20}
\newcommand*{\INFNGE}{INFN, Sezione di Genova, 16146 Genova, Italy}
\newcommand*{\INFNGEindex}{21}
\newcommand*{\INFNRO}{INFN, Sezione di Roma Tor Vergata, 00133 Rome, Italy}
\newcommand*{\INFNROindex}{22}
\newcommand*{\INFNTUR}{INFN, Sezione di Torino, 10125 Torino, Italy}
\newcommand*{\INFNTURindex}{23}
\newcommand*{\ORSAY}{Institut de Physique Nucl\'eaire, CNRS/IN2P3 and Universit\'e Paris Sud, Orsay, France}
\newcommand*{\ORSAYindex}{24}
\newcommand*{\ITEP}{Institute of Theoretical and Experimental Physics, Moscow, 117259, Russia}
\newcommand*{\ITEPindex}{25}
\newcommand*{\JMU}{James Madison University, Harrisonburg, Virginia 22807}
\newcommand*{\JMUindex}{26}
\newcommand*{\KNU}{Kyungpook National University, Daegu 41566, Republic of Korea}
\newcommand*{\KNUindex}{27}
\newcommand*{\MISS}{Mississippi State University, Mississippi State, MS 39762-5167}
\newcommand*{\MISSindex}{28}
\newcommand*{\UNH}{University of New Hampshire, Durham, New Hampshire 03824-3568}
\newcommand*{\UNHindex}{29}
\newcommand*{\NSU}{Norfolk State University, Norfolk, Virginia 23504}
\newcommand*{\NSUindex}{30}
\newcommand*{\OHIOU}{Ohio University, Athens, Ohio  45701}
\newcommand*{\OHIOUindex}{31}
\newcommand*{\ODU}{Old Dominion University, Norfolk, Virginia 23529}
\newcommand*{\ODUindex}{32}
\newcommand*{\RPI}{Rensselaer Polytechnic Institute, Troy, New York 12180-3590}
\newcommand*{\RPIindex}{33}
\newcommand*{\URICH}{University of Richmond, Richmond, Virginia 23173}
\newcommand*{\URICHindex}{34}
\newcommand*{\ROMAII}{Universita' di Roma Tor Vergata, 00133 Rome Italy}
\newcommand*{\ROMAIIindex}{35}
\newcommand*{\MSU}{Skobeltsyn Institute of Nuclear Physics, Lomonosov Moscow State University, 119234 Moscow, Russia}
\newcommand*{\MSUindex}{36}
\newcommand*{\SCAROLINA}{University of South Carolina, Columbia, South Carolina 29208}
\newcommand*{\SCAROLINAindex}{37}
\newcommand*{\TEMPLE}{Temple University,  Philadelphia, PA 19122 }
\newcommand*{\TEMPLEindex}{38}
\newcommand*{\JLAB}{Thomas Jefferson National Accelerator Facility, Newport News, Virginia 23606}
\newcommand*{\JLABindex}{39}
\newcommand*{\UTFSM}{Universidad T\'{e}cnica Federico Santa Mar\'{i}a, Casilla 110-V Valpara\'{i}so, Chile}
\newcommand*{\UTFSMindex}{40}
\newcommand*{\EDINBURGH}{Edinburgh University, Edinburgh EH9 3JZ, United Kingdom}
\newcommand*{\EDINBURGHindex}{41}
\newcommand*{\GLASGOW}{University of Glasgow, Glasgow G12 8QQ, United Kingdom}
\newcommand*{\GLASGOWindex}{42}
\newcommand*{\VT}{Virginia Tech, Blacksburg, Virginia   24061-0435}
\newcommand*{\VTindex}{43}
\newcommand*{\VIRGINIA}{University of Virginia, Charlottesville, Virginia 22901}
\newcommand*{\VIRGINIAindex}{44}
\newcommand*{\WM}{College of William and Mary, Williamsburg, Virginia 23187-8795}
\newcommand*{\WMindex}{45}
\newcommand*{\YEREVAN}{Yerevan Physics Institute, 375036 Yerevan, Armenia}
\newcommand*{\YEREVANindex}{46}
 
\author[toASU,toCUA]{P.~ Collins\fnref{Patrick}}
\author[toASU]{B.G.~ Ritchie\corref{cor1}}
\author[toASU]{M. Dugger}
\author[toBONNHI]{A.~V.~ Anisovich} 
\author[toGWUI,toJLAB]{M.~ D\"{o}ring}
\author [toBONNHI,toJLAB]{E.~ Klempt} 
\author[toBONNHI,toGATCHINA]{V.~A.~ Nikonov} 
\author[toBONNB,toBONNHI]{D.~ R\"{o}nchen} 
\author[toGWUI]{D.~ Sadasivan}
\author[toBONNHI,toGATCHINA]{A.~ Sarantsev}

% ...
\address[toASU]{\ASU} 
 \address[toBONNB]{\BONNB}
 \address[toBONNHI]{\BONNHI}
% ...
\address[toCUA]{\CUA} 
% ...
\address[toGATCHINA]{\GATCHINA}
% ...
\address[toGWUI]{\GWUI} 
% ...
\address[toJLAB]{\JLAB} 
% ...
 
\cortext[cor1]{Corresponding author}
\fntext[Patrick]{Current address: 8051 Jason Avenue, West Hills, CA 91304}
\fntext[NOWINFNGE]{Current address: INFN, Sezione di Genova,  16146 Genova, Italy }

\date{\today}

\begin{abstract}
Measurements of the linearly-polarized photon beam asymmetry $\Sigma$ 
for photoproduction from the proton 
of $\eta$ and $\eta^\prime$ mesons are reported. 
A linearly-polarized tagged photon beam produced by coherent bremsstrahlung
was incident on a cryogenic hydrogen target within the CEBAF Large Acceptance Spectrometer. 
Results are presented for the $\gamma p \to \eta p$ reaction 
for incident photon energies from 1.070 to 1.876 GeV, 
and from 1.516 to 1.836 GeV for the $\gamma p \to \eta^\prime p$ reaction. 
For $\gamma p \to \eta p$, the data reported here 
considerably extend the range of measurements to higher energies, 
and are consistent with the few previously published measurements 
for this observable near threshold.
For $\gamma p \to \eta^\prime p$, the results obtained 
are consistent with the few previously published measurements for this observable
near threshold, but also greatly expand the incident photon energy coverage for that reaction.
Initial analysis of the data reported here with the Bonn-Gatchina model 
strengthens the evidence for four nucleon resonances -- 
the $N(1895)1/2^-$, $N(1900)3/2^+$, $N(2100)1/2^+$ and $N(2120)3/2^-$ resonances --
which presently lack the ``four-star'' status in the current Particle Data Group compilation,
providing examples of how these new measurements help refine models of the photoproduction process.
\end{abstract}


\end{frontmatter}

\section{Introduction\label{section1}}

Much effort in nuclear physics at present is aimed at obtaining
a quantum-chromodynamic description of the nucleon in terms of its quark constituents. 
Our current knowledge of nucleon resonances~\cite{Crede:2013sze, Patrignani:2016,Burkert:2016kyi} 
has come from analyses of the results of experiments primarily
with $\pi N$, $\eta N$, $K \Lambda$, and $K \Sigma$ final states.
These analyses have identified (with varying degrees of certainty) 
a large number of excited states
over the past several decades
(e.g.~\cite{Tiator:2011pw,Aznauryan:2012ba,Cloet:2013jya,Bazavov:2014xya}).
Nonetheless, despite experimental efforts spanning nearly a half of a century, 
considerable ambiguity still remains about precisely which resonances indeed are present
and the details of the properties of those excitations.
The competing theoretical descriptions of the nucleon resonance spectrum predict many more
states than have been observed  (the longstanding ``missing resonance'' puzzle). 

Progress in understanding the nucleon has been difficult 
in part because of the complexity of the nucleon resonance spectrum, 
with excited states often overlapping each other in energy because of their inherently
broad width (typically 100-300 MeV). 
To better isolate specific contributions to the nucleon excitation spectrum, 
studies using the electromagnetic interaction have proven to be powerful, 
since the features of that interaction are well understood in terms of quantum electrodynamics
and since photons potentially might have large couplings to resonances 
that have escaped detection in previous analyses of reactions using pion beams. 
The reactions $\gamma p \to \eta p$ and $\gamma p \to \eta^\prime p$ 
have been seen to be quite advantageous
in probing the nucleon since 
those reactions provide an ``isospin filter'' on the nucleon resonance spectrum: 
the final states $\eta p$ and $\eta^\prime p$ 
can only be accessed in one-step decays of isospin $I$ = $\frac{1}{2}$ resonances, 
whereas data with $\pi N$ final states, 
which make up the bulk of the current world database, 
include both $I$ = $\frac{1}{2}$ and $\frac{3}{2}$ resonances.

Most published studies of the reactions $\gamma p \to \eta p$ and $\gamma p \to \eta^\prime p$ below
an incident photon energy $E_\gamma$ of 2 GeV 
have focused on measurements of the  
differential cross section~\cite{Krusche:1995nv, Dugger:2002ft, Crede:2003ax, Nakabayashi:2006ut, Bartalini:2007fg, Crede:2009zzb, Williams:2009yj, Sumihama:2009gf, McNicoll:2010qk, Plotzke:1998ua, Dugger:2005my}. 
Cross section data have helped delineate the basic features for nucleon excitations in that energy range, 
but that observable alone does not provide sufficient information
for resolving the details of the nucleon resonance spectrum. 
To gain further insight, more recent studies have turned to spin observables, 
wherein the interferences of  helicity amplitudes~\cite{Barker:1974vm,Barker:1975bp} 
provide much more detailed and stringent tests 
of the predictions arising from various theoretical models of excited nucleon states. 
For pseudoscalar meson photoproduction,
there are a total of 16 possible observables using polarized and unpolarized photons,
polarized and unpolarized proton targets, 
and measurements of the polarization of the recoiling proton following photoproduction.
As outlined in Ref.~\cite{Barker:1975bp}, in principle, 
full knowledge of all the helicity amplitudes for the process
for a particular incident photon energy $E_\gamma$ 
(or, alternately, center-of-mass energy $W$)
can be obtained 
with a judicious choice of 
a subset of 8 of the 16 possible observables,
thereby providing a so-called ``complete" measurement of the helicity amplitudes.
However, when experimental uncertainties are considered,
many ambiguities usually remain even with such a 
choice~\cite{Sandorfi:2013gra,Ireland:2010bi,Nys:2016uel}.
Consequently, increasing accuracy in any theoretical description demands  
extending the dataset on all observables as much as possible.

As part of the effort to gain a more complete dataset of measurements,
the work reported here provides data on the photon beam asymmetry observable $\Sigma$ 
for the reactions $\gamma p \to \eta p$ and $\gamma p \to \eta^\prime p$. 
The photon beam asymmetry $\Sigma$ is defined in the center-of-mass frame as 
\begin{equation} \label{eq:Sigmadef}
	\frac{d\sigma}{d\Omega} = 
		\frac{d\sigma_0}{d\Omega} \left[1 - P_\gamma \Sigma \cos \{ 2 \left ( \varphi-\alpha \right) \} \right],
\end{equation}
where $\frac{d\sigma}{d\Omega}$ is the differential cross section for the reaction
using a polarized photon beam, 
 $\frac{d\sigma_0}{d\Omega}$ is the {\em{unpolarized}} differential cross section, 
$P_\gamma$ is the degree of linear polarization of the photon beam, 
$\varphi$ is the azimuthal angle of the photoproduced meson
relative to a plane parallel to the floor in the laboratory frame,
and $\alpha$ is the azimuthal angle between the photon beam polarization plane
and the laboratory floor plane.  
The beam asymmetry $\Sigma$ is particularly powerful for testing resonance descriptions of the nucleon 
since this observable may be written as $2 Re(S_1 ^\ast S_2 - {\cal{N}} D^\ast )$, 
where $S_1$ and $S_2$ are the $s$-channel single-flip helicity amplitudes 
and $\cal{N}$ and $D$ are the no-flip and double-flip $s$-channel helicity amplitudes,
respectively; 
thus, measurements of $\Sigma$ help isolate those various components
through interference effects \cite{Barker:1974vm}. 
While there have been several measurements of $\Sigma$ for $\gamma p \to \eta p$~\cite{Bartalini:2007fg, Elsner:2007hm, Vartapetian:1980cn},
only one previous publication has reported 
$\Sigma$ data for $\gamma p \to \eta^\prime p$~\cite{Sandri:2014nqz},
in that case providing $\Sigma$ for two energies near the $\eta^\prime p$ threshold.
The authors of Ref.~\cite{Sandri:2014nqz} drew attention to an
intriguing $\sin^2 \theta_{c.m.}\cos \theta_{c.m.}$ angular dependence near
threshold for $\Sigma$ that was not reproduced by the theories discussed in that work,
and noted that such a feature would be suggestive of 
interference between either $P$- and $D$-waves or $S$- and $F$-waves;
if true, such behavior would require at least one additional resonance beyond
the four resonances that have been suggested to be important
near threshold ($N(1720)3/2^+$, $N(1925)1/2^-$,
$N(2130)1/2^+$, and $N(2050)3/2^+)$~\cite{Huang:2012xj}.

The results presented here for the photon beam asymmetry $\Sigma$ 
provide a check on prior measurements for both these reactions,
but also extend the measurements of $\Sigma$ to considerably higher energies than previously reported,
thereby providing access with this observable to the details of higher-lying resonances. 
To provide an indication of the utility of these new $\Sigma$ data, 
comparisons are provided with a number of models,
and initial investigations are presented using two approaches that
take advantage of the newest data on these reactions. 

\section{Experiment}
The experiment was conducted in Hall B 
at the Thomas Jefferson National Accelerator Facility (Jefferson Lab) 
during the ``g8b'' running period, 
which also provided the data from which photon beam asymmetries for 
$\pi^+$ and $\pi^0$ photoproduction on the proton
were extracted~\cite{Dugger:2013crn}, 
as well as data related to strangeness-related photoproduction on the proton~\cite{Paterson:2016vmc}.
Full details on the experimental conditions for that running period may be found in 
those publications, but a summary is provided here. 

A linearly-polarized photon beam was generated 
by coherent bremsstrahlung~\cite{Bilokon:1982wu} using
a 4.55 GeV electron beam and an oriented 50~$\mu$m-thick diamond.
The coherent bremsstrahlung process results in 
intensity enhancements in the photon spectrum above
the normal bremsstrahlung spectrum due to 
momentum transfer from the scattered electron 
to the lattice planes within the diamond; 
significant linear polarization enhancement occurs in
those intensity enhancement peaks. 
The photon energy where the intensity enhancement is greatest is called the coherent peak. 
Adjusting the orientation of the diamond controls the photon polarization plane 
as well as the coherent peak for producing polarized photons.
Energy, timing, and polarization information for the photon beam 
were provided by the Hall B photon tagger~\cite{Sober:2000we},
and the degree of photon beam polarization during each portion of the data collection period
was estimated via a bremsstrahlung calculation
using knowledge of the diamond orientation and the
degree of photon beam collimation~\cite{Livingston:2011}. 

The photon beam was incident 
on a 40-cm-long cryogenic liquid hydrogen target
placed 20 cm upstream from the center 
of the CEBAF Large Acceptance Spectrometer (CLAS)~\cite{Mecking:2003zu},
composed of six identical charged particle detectors installed in a toroidal magnetic field. 
The principal CLAS subsystems used here were: 
the drift chamber system 
for tracking charged particles~\cite{Mestayer:2000we}, 
with three multi-layer drift chambers in each sector of CLAS,
yielding a total of approximately 35,000 individually instrumented hexagonal drift cells; 
a scintillator-based time-of-flight (TOF) system~\cite{Smith:1999ii}, 
with 57 elements per sector; 
and a ``start counter'' plastic scintillator array, with six elements per sector,
which determined when charged particles passed from the target into the detection region~\cite{Sharabian:2005kq}. 

To determine $\Sigma$, 
Eq.~(\ref{eq:Sigmadef}) may be recast 
based on the orientation of the plane of polarization for the electric field $\vec{E}$
of the photon beam relative to the lab floor:
\begin{itemize}
\item[(a)] ``perpendicular beam'' polarization ($\vec{E} \perp$ to lab floor),
\begin{equation}\label{eq:Sigma_perp}
\sigma_\perp(\theta,\varphi) \equiv \frac{d\sigma_\perp}{d\Omega}(\theta,\varphi) 
=\frac{d\sigma_0}{d\Omega}(\theta) [1+P_\perp \Sigma \cos 2 \varphi] \ ,
\end{equation}
\item[(b)] ``parallel beam'' polarization ($\vec{E} \para$ to lab floor),
\begin{equation}\label{eq:Sigma_para}
\sigma_\para(\theta,\varphi) \equiv \frac{d\sigma_\para}{d\Omega}(\theta,\varphi)=\frac{d\sigma_0}{d\Omega}(\theta) [1-P_\para \Sigma \cos 2\varphi] \ ,
\end{equation}
\end{itemize}
where $P_\perp$ and $P_\para$ denote the degrees of photon beam polarization 
for the perpendicular and parallel polarization orientations, respectively;
$\theta_{c.m.}$ denotes the meson polar scattering angle in the center-of-mass system
(hereafter $\theta_{c.m.}$). 
With Eqs.~(\ref{eq:Sigma_perp}) and~(\ref{eq:Sigma_para})$, \Sigma$ then may be written as
\begin{equation}\label{eq:Sigma}
\Sigma=\frac{\sigma_\perp - \sigma_\para}{\sigma_\perp + \sigma_\para} \ .
\end{equation}
The g8b running period was divided into intervals 
with different coherent peak energies nominally set to 1.3, 1.5, 1.7, and 1.9 GeV. 
These intervals were further subdivided into periods with parallel or perpendicular beam orientation.

\subsection{Particle and event identification}

For each charged particle detected in CLAS,
the accelerator radio-frequency (RF) timing 
and information from the start counter array, TOF subsystem, and drift chambers 
were used to determine each charged particle's type and four-momentum, as well as
to identify which tagged photon gave rise to the reaction generating that particle.  
The vertex time (i.e. the time when the reaction took place in the cryogenic target) 
for each event was established 
using the time difference 
between the time of passage for that particle 
through the start counter (at the entrance to the drift chamber region) 
and the corresponding time of passage 
through a counter in the TOF array (as the particle exited the drift chamber region).
This vertex time was then used to identify which tagged photon gave rise to the reaction
that produced that particular charged particle. 
Once the particular tagged photon was identified for the event,
the RF-corrected photon vertex time and TOF information were used to 
identify the type of charged particle and to make sure that all charged particles
assumed to be in a particular event
were indeed associated with the same photon and event. 

The drift chambers provided trajectory information on each scattered particle, 
and the combination of timing information 
and trajectory information 
yielded a velocity and momentum determination for each charged particle. 
Particle identification then was performed using an algorithm that compared the 
CLAS-measured momentum of the particle whose identity was to be determined with 
expected values of $\beta$ for the possible identities for that particle~\cite{GPID}.
Each possible identity was tested by comparing the expected value of $\beta$ 
for a given particle type to the CLAS-measured value of $\beta$ 
determined by CLAS tracking and time-of-flight information. 
The particle was then assigned the identity that provided 
the closest expected value of $\beta$ to the empirically measured value of $\beta$. 
The performance of this particle identification technique is illustrated in
Fig.~1 of Ref.~\cite{Dugger:2013crn}.

A correction due to energy loss in the target and detector materials was performed
for each charged particle identified, 
with the 4-vector values adjusted accordingly~\cite{Pasyuk:2007a}. 
The tracks and the event as a whole were associated with a particular beam photon based 
on the consistency of timing information from the photon tagger and the projected vertex timing. 
Momentum corrections for tracks in CLAS were then determined 
by demanding four-momentum conservation in a kinematic fit of
a large sample of $\gamma p \rightarrow \pi^+\pi^- p$ events seen in the spectrometer
where all three final-state particles were detected.
To avoid ambiguity, only events with particles matching exactly one beam photon were kept.
The energy calibration of the photon tagger was determined as described in Ref.~\cite{CLASg8b},
resulting in a photon energy resolution typically better than $\pm$ 0.1 MeV, 
and always better than $\pm$ 0.5 MeV. 

The scattering angle and momentum for the proton recoiling from meson photoproduction 
were used to calculate a missing mass $M_X$
from a two-body final state 
based on the assumption that the reaction observed was $\gamma p \to pX$, 
where $X$ was the other body in the two-body final state.

\subsection{Reconstruction of $\eta$ and $\eta^\prime$ mesons}
\begin{figure}%
   \centering
   \subfloat{{\includegraphics[width=.45\linewidth]{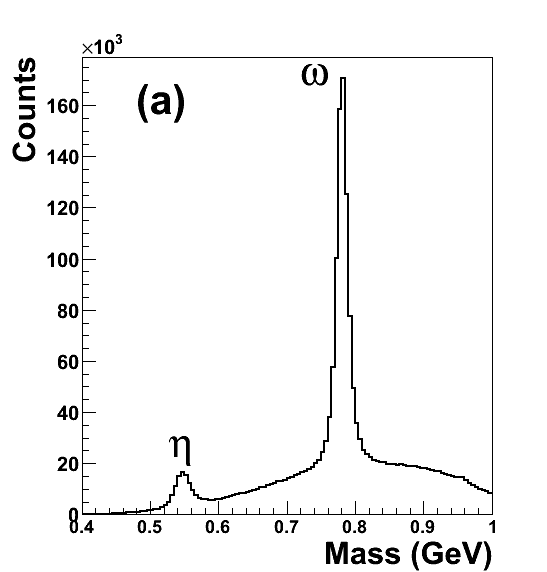}}}%
   \qquad
   \subfloat{{\includegraphics[width=.45\linewidth]{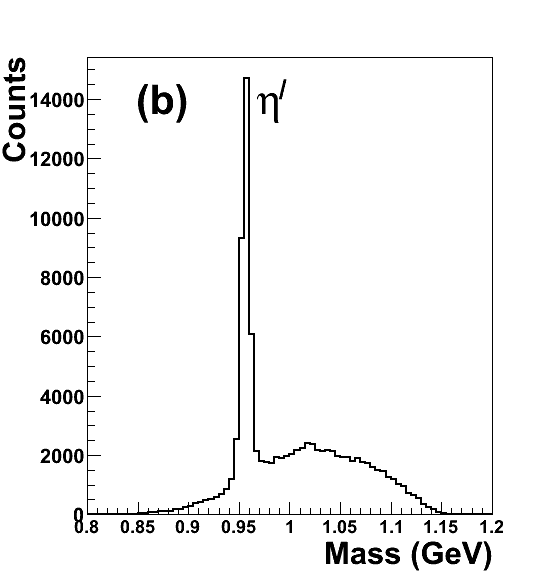}}}%
\caption{Missing mass spectra for $\gamma p \rightarrow p + X$
summed over all coherent peak settings, with
multi-pion background reduced by requiring
detection in CLAS of a proton, $\pi^+$, and $\pi^-$. 
The peaks associated with the $\eta$ and $\eta^\prime$ photoproduced mesons
are indicated. 
(a) Isolation of $\eta$ photoproduction using $p\pi^+\pi^-(\pi^0)$. 
(b) Isolation of $\eta^\prime$ photoproduction using $p\pi^+\pi^-(\eta)$.
\label{eta-prime-MM}}
\end{figure}

The missing mass spectra constructed in this fashion possessed  
considerable background from multi-pion photoproduction under the 
peaks asssociated with the $\eta$ and $\eta^\prime$ mesons. 
For this reason, detection of the decay products for the $\eta$ and $\eta^\prime$ mesons 
was incorporated to reduce this background.
The $\eta$ decays to the state $\pi^+\pi^-\pi^0$ with a branching ratio of 22.9\%~\cite{Patrignani:2016}, 
while the $\eta^\prime$ decays to the state $\pi^+\pi^-\eta$ 
with a branching ratio of 42.9\%~\cite{Patrignani:2016}. 
The charged pions resulting from these decays were detected in CLAS, 
and the remaining neutral mesons were then identified with the missing mass technique. 
Once final states with appropriate decay products were identified, 
the reactions $\gamma p \to \eta p$ and $\gamma p \to \eta^\prime p$ 
were then analyzed.
Examples of the performance of this technique 
for $\eta$ and $\eta^\prime$ are seen in Fig.~\ref{eta-prime-MM}.
(The prominent $\omega$ meson peak seen in Fig.~\ref{eta-prime-MM}(a)
also permitted measurements
of the $\Sigma$ observable for $\omega$ photoproduction, 
which will form the subject of a forthcoming publication.)

A Fourier-moment analysis technique was used to extract $\Sigma$
as used previously for the $\pi^0$ and $\pi^+$ photon beam asymmetry measurements
reported in Ref.~\cite{Dugger:2013crn}. 
Cosine-$\varphi$-moment histograms were constructed 
by taking each event in the $\gamma p \to \eta p$ 
or $\gamma p \to \eta^\prime p$ missing mass histograms 
and weighting that event by the value of $\cos n\varphi$ corresponding to that event. 
With this approach, events within a particular $\cos \theta_{c.m.}$ bin 
for any value of $\varphi$ 
are combined simultaneously to determine $\Sigma$. 
Applying this Fourier moment method to $\Sigma$, 
the resulting equation for the beam asymmetry may be written
\begin{equation}\label{eq:Sigma_moments}
\Sigma=\frac{\tilde{Y}_{\perp 2} - \tilde{Y}_{\para 2}}
{\frac{P_\para}{2}(\tilde{Y}_{\perp 0} + \tilde{Y}_{\perp 4}) + \frac{P_\perp}{2}(\tilde{Y}_{\para 0} + \tilde{Y}_{\para 4})} \ , 
\end{equation}
where  $\tilde{Y}_{\perp n}$($\tilde{Y}_{\para n}$) is the normalized meson yield 
for a perpendicular (parallel) photon beam, with each event weighted 
according to the Fourier moment $\cos  n \varphi $, and $P_\perp$( $P_\para$) is the degree of photon polarization.

\subsection{Kinematic bins}

The data were sorted into kinematic bins based on photon energy $E_\gamma$ 
and center-of-mass polar angle $\cos\theta_{c.m.}$ for the photoproduced meson.
The photon energy widths of these kinematic bins were chosen
 to minimize statistical uncertainties for the extracted beam asymmetries while
providing the best center-of-mass energy $W$ resolution possible for the nucleon resonance spectrum.

With those factors in mind, and
selecting specific groups of physical counters on the photon tagger focal plane,
the $\eta$ photoproduction data were analyzed  in 
27-\unit{MeV}-wide $E_\gamma$ bins 
and 0.2-wide center-of-mass $\eta$ polar angle $\cos\theta_{c.m.}$ bins,
except for the 1.9-\unit{GeV} coherent peak, where the number of events was sufficiently low that 
the width of the $E_\gamma$ bins was increased to 54 \unit{MeV}. 
Due to the much smaller cross section for $\gamma p \rightarrow \eta^\prime p$, 
the same considerations led to an  
$E_\gamma$ bin width of 54 \unit{MeV} for all coherent peak settings for the $\eta^\prime$ results. 

\subsection{Uncertainties in extracted $\Sigma$ values}
As would be expected from 
the expression for the beam asymmetry $\Sigma$ in Eqs. (\ref{eq:Sigma})
and (\ref{eq:Sigma_moments}),
experimental quantities related to target density,
detector acceptance, and detection efficiency cancel in such a ratio,
so those quantities did not contribute to systematic or statistical uncertainties.
The statistical uncertainty in the relative normalization of the photon beam flux for 
the different coherent peak settings and 
polarization orientations was much less than 1\%,
contributing negligibly to the statistical uncertainty
for $\Sigma$ at any incident photon energy.
Overall, the statistical uncertainties for $\Sigma$ were driven
by the uncertainties in the yield,
though the use of the Fourier moment method 
as in Eq.~\ref{eq:Sigma_moments}
requires
careful propagation of uncertainties in the various moments,
as well as the correlations between parts of the ratio for $\Sigma$,
as outlined in Ref.~\cite{Dugger:2013crn}. 
Statistical uncertainties varied markedly
from point to point owing to
the underlying variations in the photoproduction cross sections and $\Sigma$, but
the average absolute statistical uncertainty $\langle{\Delta\Sigma}\rangle$ in $\Sigma$ was on the order of
$\langle{\Delta\Sigma}\rangle=\pm$ 0.15 for both $\gamma p \to \eta p$
and $\gamma p \to \eta^\prime p$.

Systematic uncertainties for $\Sigma$ were driven 
by the systematics of the polarization estimation and relative normalization. 
By analyzing measurements at different coherent peak settings 
but where photon energies were the same,
the systematic uncertainty in the photon polarization 
for a particular polarization orientation was found to be 4\%,
as reported in Ref.~\cite{Dugger:2013crn}.
When combining data taken with the two different polarization orientations, 
adding those contributions in quadrature resulted in an 
estimated systematic uncertainty in $\Sigma$ of 6\%,
as given in Ref.~\cite{Dugger:2013crn}.

\section{Results}
\begin{figure*}
\centerline{\includegraphics[width=7in]{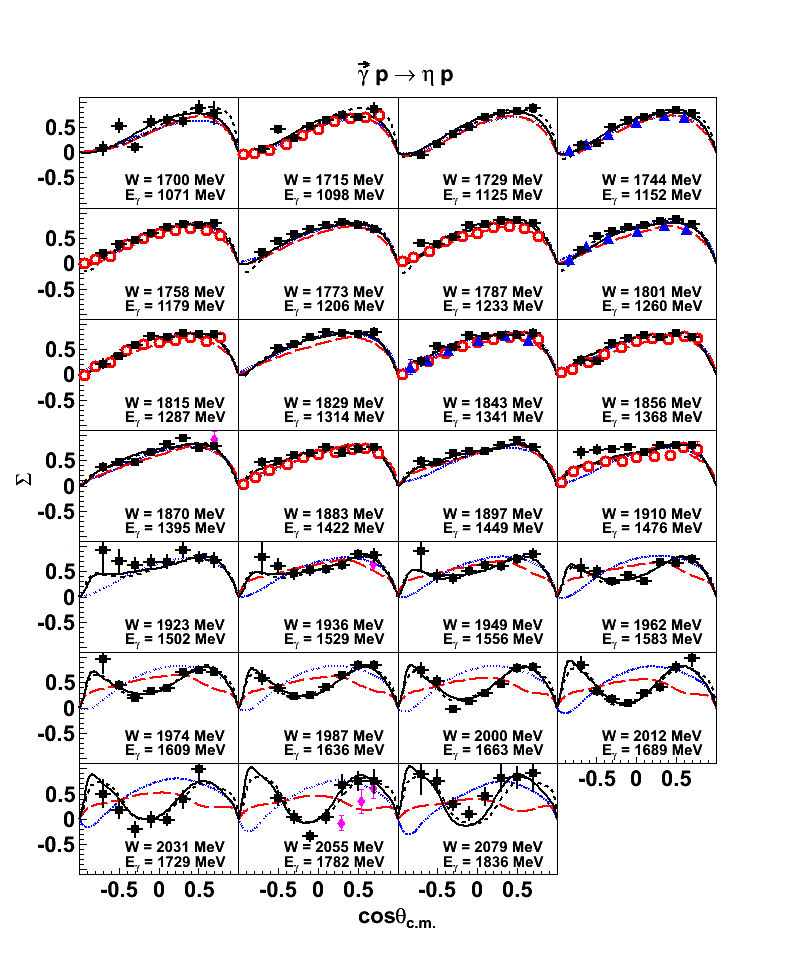}}
\caption{(Color online) The beam asymmetry $\Sigma$ 
as a function of $\cos\theta_{c.m.}$ for the reaction $\gamma p \to \eta p$
at incident photon energies $E_\gamma$ from 1.071 GeV  
($W$ = 1.700 GeV) to 1.836 GeV  ($W$ = 2.079 GeV). 
The data reported here are shown as black filled squares. 
Previously published results from 
Refs.~\protect{\cite{Elsner:2007hm,Bartalini:2007fg,Vartapetian:1980cn}}
are shown as (blue) filled triangles. (red) open circles, and (pink) diamonds, respectively.  
The (blue) dotted lines indicate SAID predictions~\protect{\cite{Workman:2012hx}}, 
while predictions from the ETA-MAID model~{\cite{Kashevarov:2016owq}} 
indicated by the (red) long-dashed lines.
Results from new fits with the J\"{u}lich-Bonn model~\protect{\cite{Ronchen:2015vfa}} as discussed in the text
are shown 
with (black solid lines) and without (black short-dashed lines) the inclusion of a $N(1900)3/2^+$ resonance. 
 \label{eta6x5}}
\end{figure*}

\begin{figure}[ht]
\centerline{\includegraphics[width=3.75in]{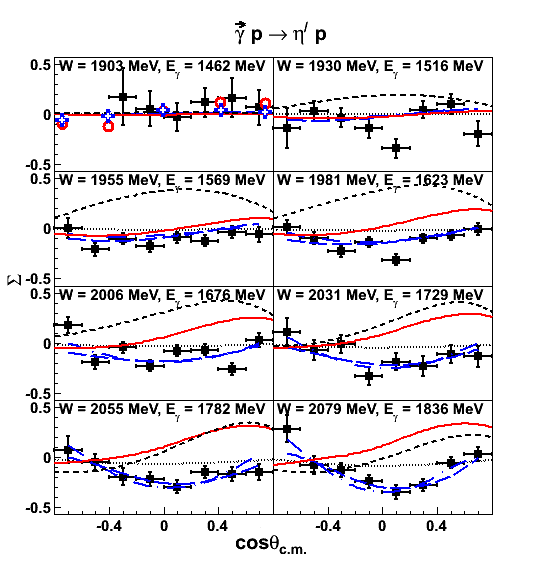}}
\caption{ (Color online) Beam asymmetries as a function of $\cos\theta_{c.m.}$ 
for the reaction $\gamma p \to \eta^\prime p$ at incident photon energies  $E_\gamma$  from 1.462 GeV  
($W=$1.903 GeV) (top left) to 1.836 GeV  ($W=$2.079 GeV) (bottom right). 
CLAS data are shown by black squares.
Prior measurements from GRAAL~\protect{\cite{Sandri:2014nqz}}
are shown as open symbols,
with (red) circles for measurements at $E_\gamma$=1.461 GeV and
(blue) crosses for measurements at $E_\gamma$=1.480 GeV.
Also shown are predictions from the SM05 solution of SAID~\protect{\cite{Workman:2012hx}} 
(indicated by dotted blue lines), 
from ETA-MAID~\protect{\cite{Chiang:2002vq}} (solid red lines), 
and from Nakayama and Haberzettl (model 4 in \protect{\cite{Nakayama:2005ts}}, dashed black lines). 
Two new fits discussed in the text using the Bonn-Gatchina model that including
the data reported here are also shown (long-dashed and dashed-dotted blue lines). 
\label{etaPrime4x2}}
\end{figure}

The photon beam asymmetry $\Sigma$ results obtained here are shown in Figs.~\ref{eta6x5}  
and \ref{etaPrime4x2} for $\eta$ and $\eta^\prime$, respectively. 
A total of 266 data points  for $\Sigma$ distributed over 27 bins 
in incident photon energy $E_\gamma$ for $\gamma p \to \eta p$ 
were obtained, and 62 data points for $\Sigma$ 
in 8 bins in $E_\gamma$ for $\gamma p \to \eta^\prime p$.

Also shown in Fig.~\ref{eta6x5} are $\Sigma$ results 
for $\gamma p \to \eta p$ from 
Refs.~\cite{Elsner:2007hm,Bartalini:2007fg,Vartapetian:1980cn}
at energies close to those for which data are reported here.
As seen in that figure, the angular dependence observed in this work for $\Sigma$
is comparable to that seen in Refs.~\cite{Elsner:2007hm,Bartalini:2007fg}, 
and the prior results and the results reported here generally agree in magnitude 
within statistical and systematic uncertainties. 
Nonetheless, uncertainties aside, the results of Ref.~\cite{Bartalini:2007fg}, while similar in shape,
are systematically smaller in magnitude than the results reported here, and
the results at $E_\gamma$=1.476 GeV from that reference also disagree in
shape with those reported here. 
Careful inspection of the data reported here did not reveal 
any specific sources for these differences.
For all but the $E_\gamma$=1.476 GeV results,
the agreement in shape suggests the source of the disagreement could be
attributable to the polarization estimate in either or both cases.
The disagreement at 1.476 GeV, however, suggests additional sources
beyond the estimate of the photon beam polarization
may be responsible for the discrepancies observed. 
The data from Ref.~\cite{Vartapetian:1980cn}
at $W$=2.055 GeV
disagree sharply beyond uncertainties in both magnitude and shape 
with the data provided here. 
It is unclear why this disagreement arises, but, given the general
agreement within uncertainties between the data reported here and the more recent
studies in Refs.~\cite{Elsner:2007hm,Bartalini:2007fg},
except as noted above,
a problem with the earlier data might exist. 
More $\Sigma$ data for $\gamma p \to \eta p$
near $W$=2.0 GeV are needed to clarify the situation. 

The recent $\Sigma$ results from GRAAL
for $\gamma p \to \eta^\prime p$~\cite{Sandri:2014nqz},
which represent the only other measurements of $\Sigma$ for that reaction,
are compared with the results obtained here
in Fig.~\ref{etaPrime4x2}. 
The two data sets are consistent with each other
within our comparatively large uncertainties for
the lowest of the 8 energy bins reported here.

\section{Discussion}
As noted in the Introduction, these data can help test and refine
theoretical descriptions of the reaction process via the participation
of various nucleon resonances, particularly when coupled with
other observables for one or more photoproduction reactions. 
In turn, those descriptions can motivate and clarify QCD-based
descriptions of the nucleon. For the present discussion, 
we consider each reaction in turn, 
presenting comparisons with previous predictions for $\Sigma$.
We also provide initial results of an investigation of each reaction using these new data
through new fits with two existing models. 
For the $\eta$, initial results based on the J\"{u}lich-Bonn approach~\cite{Ronchen:2015vfa} are
presented, while the $\eta^\prime$ results are discussed in the
context of the Bonn-Gatchina model~\cite{Anisovich:2004zz,Anisovich:2006bc,Denisenko:2016ugz}. 

\subsection{Photon beam asymmetry $\Sigma$ for $\gamma p \to \eta p$}

The results from this work for $\Sigma$ in $\gamma p \to \eta p$ are compared to predictions from
 SAID~\cite{Workman:2012hx} and ETA-MAID~\cite{Kashevarov:2016owq}
 in Fig.~\ref{eta6x5}.
Those predictions provide satisfactory descriptions of the observable 
below $E_\gamma$=1.5~GeV ($W$=1.9 GeV),
as might be expected where prior data exist.
For higher energies, however, the angular dependence of $\Sigma$ is not reproduced satisfactorily, 
particularly around $W$=2.0 GeV. 

As an initial application of this new dataset, 
new predictions for $\Sigma$ for $\gamma p \to \eta p$ 
using the J\"{u}lich-Bonn model~\cite{Ronchen:2015vfa} 
were developed by fitting to these new data shown in Fig.~\ref{eta6x5}, 
and those new predictions are shown in that figure. 
The J\"{u}lich-Bonn model uses 
a dynamical coupled-channels approach, 
wherein the hadronic scattering amplitude
is constructed with a potential generated 
from an effective SU(3) Lagrangian 
through time-ordered perturbation theory, 
with unitarity and analyticity automatically preserved.
The new predictions used the same parameters 
incorporated for the fit discussed in Ref.~\cite{Senderovich:2015lek}, 
and included the results for the $E$ observable of that work. 
The fit also simultaneously incorporated the world databases 
for the pion-induced production of $\eta N$, $K\Lambda$, and $K\Sigma$ final states~\cite{Ronchen:2012eg} 
and the partial-wave solution of the SAID group~\cite{Workman:2012hx} for elastic $\pi N$ scattering. 
Most published data on observables for pion and $\eta$ photoproduction on the proton up to $W \approx$ 2.3 GeV 
were incorporated into the database for fitting~\cite{Ronchen:2015vfa, Ronchen:2014cna}, 
including the recent MAMI results on $T$ and $F$  for $\eta$ photoproduction~\cite{Akondi:2014ttg}, 
for a total database of nearly 30,000 points.

In the refits with the new $\Sigma$ data, 
both the generalized variance and multi-collinearity of helicity couplings improved significantly, 
demonstrating that this new data help to refine the values 
for the electromagnetic properties of resonances on an absolute scale, 
and also to reduce the correlations between resonances.
The various helicity couplings remained relatively stable for most resonances
 after inclusion of the new $\Sigma$ data in the fit, 
 but helicity couplings for 
the $N(1720)3/2^+$ and the $N(1900)3/2^+$  
changed noticeably;
for example, the amplitude $|A^{1/2}|$ for the  $N(1720)3/2^+$ resonance fell by nearly half. 
Previously, the $N(1900)3/2^+$ was found to be important in analyses by the Bonn-Gatchina group of 
$K\Lambda$ and $K\Sigma$ photoproduction~\cite{Anisovich:2012ct}.

To illustrate the effect of the sizeable changes in the parameters for the $N(1900)3/2^+$ state, 
two sets of predictions are shown in Fig.~\ref{eta6x5}, 
where the sole difference in the model is the presence of a $N(1900)3/2^+$ contribution. 
As seen in the figure, both predictions for $\Sigma$ perform comparably well
($\chi^2$/point $\approx$ 1.4) throughout the $W$ range measured here. 
Thus, this comparison suggests the parameters of the $N(1900)3/2^+$ 
are not particularly well constrained in the $\eta$ photoproduction reaction process 
by the $\Sigma$ observable, 
but further investigation is warranted 
to better understand the interdependencies of the resonance parameters 
within this model so as to further constrain the resonance description of the reaction.
Such a study is underway and will be published subsequently~\cite{JuBoPC2016}.

\subsection{Photon beam asymmetry $\Sigma$ for $\gamma p \to \eta^\prime p$}
The data obtained here for $\Sigma$ for $\gamma p \to \eta^\prime p$ are compared
to several sets of predictions in Fig.~\ref{etaPrime4x2}, as well as the recent
data from GRAAL~\cite{Sandri:2014nqz}.
The predictions include SAID~\cite{Workman:2012hx},
Nakayama and Haberzettl~\cite{Nakayama:2005ts},   
and ETA-PRIME-MAID~\cite{Chiang:2002vq}.
In contrast to the situation for $\gamma p \to \eta p$, 
none of these predictions provides a
satisfactory description of the $\gamma p \to \eta^\prime p$ data; indeed,
the predictions generally have the wrong sign for $\Sigma$.

As an initial investigation of the data for this observable,
these data have been incorporated in a new fit using the Bonn-Gatchina
modified $K$-matrix approach~\cite{Anisovich:2004zz,Anisovich:2006bc,Denisenko:2016ugz}, 
combining contributions
from nucleon resonances and from non-resonant background
processes. Additionally, a phenomenologically``Regge-ized'' 
amplitude is used to describe vector meson exchange in the $t$ channel
by taking advantage, in part, of Reggeon-resonance 
duality~\cite{Veneziano:1974dr,Anisovich:2004zz,Anisovich:2006bc}.

Two solutions, equally good at describing the data (in terms of
$\chi^2$/point$\approx$1.5), were obtained, and both
are shown in Fig.~\ref{etaPrime4x2}.
(We note that these solutions were also simultaneously used to
fit pion and $\eta$ photoproduction data, as discussed in the
prior publications for the Bonn-Gatchina model.) 
The resonances found to be important in these solutions were
the same as in the prior work~\cite{Anisovich:2004zz,Anisovich:2006bc,Denisenko:2016ugz}, 
but the strengths of the contributions were considerably different.
The resonances found to participate also differed
from those found in Ref.~\cite{Huang:2012xj}.
Notably, both new solutions indicate the dominance of the $N(1895)1/2^-$
resonance near threshold, even though this resonance is given only 
``two-star'' status in the most recent summary of the Particle Data Group (PDG)~\cite{Patrignani:2016}.
Both solutions indicate the presence of the $N(2100)1/2^+$ and $N(2120)3/2^-$, 
rated with ``one-star'' and ``two-star'' overall status, respectively, in the most recent PDG summary.
However, in contrast to the preceding discussion of the $\eta$ asymmetry,
both solutions require a strong contribution from the $N(1900)3/2^+$
resonance to explain $\Sigma$ for $\eta^\prime$ photoproduction. 

We noted above that the GRAAL measurements~\cite{Sandri:2014nqz} 
observed that  
$\Sigma \sim \sin^2{\theta_{c.m.}}\cos{\theta_{c.m.}}$
near threshold, 
which might indicate 
presence of a $D$- or $F$-wave
resonance (or both).
Neither solution reproduced that behavior, 
though no such contribution was explicitly included
during this initial investigation.  

Thus, analysis of the data reported here with the Bonn-Gatchina model 
strengthens the evidence for four nucleon resonances -- 
the $N(1895)1/2^-$, $N(1900)3/2^+$, $N(2100)1/2^+$ and $N(2120)3/2^-$ resonances --
which presently lack the ``four-star'' status in the PDG. 
Further investigations, as well as the need for 
additional resonances beyond those discussed in the Introduction,
are underway and will be published subsequently~\cite{BoGaPC2016}.

\section{Conclusion}
In conclusion, extensive measurements of the photon beam asymmetry $\Sigma$
 for $\gamma p \to \eta p$ and $\gamma p \to \eta^\prime p$ are reported here. 
The new data significantly expand the range of photon energies 
for which this observable has been measured for $\eta$ photoproduction on the proton,
and represent the first measurements of $\Sigma$ for $\eta^\prime$ photoproduction on the proton
for photon energies considerably above threshold.
In the case of $\eta$ photoproduction, the new data compare favorably with
two previously published studies at lower energies,
but disagree sharply with the few results obtained  near $W$=2.06 GeV;
further data on $\Sigma$ at that energy would be helpful to clarify the situation there. 
 
Investigations of $\Sigma$  for $\eta$ photoproduction on the proton using the J\"{u}lich-Bonn
approach found that, when the new data reported here are considered,
the helicity couplings for the $N(1720)3/2^+$ and $N(1900)3/2^+$ states were changed significantly, 
but that the evidence for the latter resonance in the $\Sigma$ data was inconclusive for $\eta$ photoproduction. 
By contrast, studies for $\eta^\prime$ photoproduction with the Bonn-Gatchina model 
found the $N(1900)3/2^+$ to be very important. 
Taken together, the analyses provide evidence to strengthen the case for 
the $N(1895)1/2^-$, $N(1900)3/2^+$, $N(2100)1/2^+$ and $N(2120)3/2^-$ resonances.
Further studies with these two approaches are underway,  
but these initial investigations underscore the importance of using spin observables
in multiple reaction channels to elucidate the nucleon resonance spectrum,
as fits using cross section data alone or a single channel can be ambiguous.
Future measurements of other polarization observables 
(e.g., $T$, $E$, $F$, $G$, and $H$~\cite{Barker:1974vm,Sandorfi:2013gra,Ireland:2010bi,Nys:2016uel}) 
for the reactions studied here,
including measurements with so-called ``frozen spin" targets, 
can be coupled with similar measurements for other meson production reactions
to more stringently test and constrain models of the photoproduction process,
as the discussion above indicates.
Such combined analyses 
will result in further significant improvements 
in our understanding of the quark structure of the nucleon.

\section{Acknowledgements}
The authors gratefully acknowledge the assistance of the Jefferson Lab staff, and
express appreciation for resources provided by the J\"{u}lich Supercomputing Centre at the 
Forschungszentrum  J\"{u}lich. 

This work was supported by
the U.~S.~ National Science Foundation, 
the U.~S.~ Department of Energy,
the Chilean Comisi\'on Nacional de Investigaci\'on Cient\'ifica y Tecnol\'ogica (CONICYT), 
the Deutsche Forschungsgemeinschaft, 
the French Centre National de la Recherche Scientifique,
the French Commissariat \`{a} l'Energie Atomique,
the Italian Istituto Nazionale di Fisica Nucleare,
the National Natural Science Foundation of China,
the National Research Foundation of Korea,
the Scottish Universities Physics Alliance (SUPA),
and the U.~K.~ Science and Technology Facilities Council.
This material is based
upon work supported by the U.S. Department of Energy, Office of Science,
Office of Nuclear Physics under contract DE-AC05-06OR23177.
The Southeastern Universities Research Association (SURA) operates the
Thomas Jefferson National Accelerator Facility for the United States
Department of Energy under contract DE-AC05-84ER40150.

\section{References}

\bibliography{ref-tilda}{}

\end{document}